# Thermal Control System to Easily Cool the GAPS Balloon-borne Instrument on the Ground

By Hideyuki F\ukE,[1,3] Shun Okazaki,[2] Akiko Kawachi,[3] Shohei Kobayashi,[3] Masayoshi Kozai,[4] Hiroyuki Ogawa,[1] Masaru Saijo,[5] Shuto Takeuchi,[6] and Kakeru Tokunaga[1]

[1] *Institute of Space and Astronautical Science (ISAS), JAXA, Sagamihara, Japan*
[2] *Research and Development Directorate, JAXA, Tsukuba, Japan*
[3] *Graduate School of Science, Tokai University, Hiratsuka, Japan*
[4] *Joint Support-Center for Data Science Research, Research Organization of Information and Systems (ROIS), Tachikawa, Japan*
[5] *Research and Development Directorate, JAXA, Sagamihara, Japan*
[6] *Graduate School of Science and Engineering, Aoyama Gakuin University, Sagamihara, Japan*

This study developed a novel thermal control system to cool detectors of the General AntiParticle Spectrometer (GAPS) before its flights. GAPS is a balloon-borne cosmic-ray observation experiment. In its payload, GAPS contains over 1000 silicon detectors that must be cooled below −40°C. All detectors are thermally coupled to a unique heat-pipe system (HPS) that transfers heat from the detectors to a radiator. The radiator is designed to be cooled below −50°C during the flight by exposure to space. The pre-flight state of the detectors is checked on the ground at 1 atm and ambient room temperature, but the radiator cannot be similarly cooled. The authors have developed a ground cooling system (GCS) to chill the detectors for ground testing. The GCS consists of a cold plate, a chiller, and insulating foam. The cold plate is designed to be attached to the radiator and cooled by a coolant pumped by the chiller. The payload configuration, including the HPS, can be the same as that of the flight. The GCS design was validated by thermal tests using a scale model. The GCS design is simple and provides a practical guideline, including a simple estimation of appropriate thermal insulation thickness, which can be easily adapted to other applications.

**Nomenclature**

| | | |
|---|---|---|
| $C_V$ | : | specific heat at constant volume, J/kg K |
| $D$ | : | diameter, m |
| $k$ | : | thermal conductivity, W/m K |
| $L$ | : | length, m |
| $L_H$ | : | latent heat, J/kg |
| $\dot{m}$ | : | mass flow rate, kg/s |
| $P$ | : | pressure, Pa |
| $Q$ | : | amount of heat, W |
| $Re$ | : | Reynolds number |
| $S$ | : | surface area, m$^2$ |
| $T$ | : | temperature, K |
| $u$ | : | velocity, m/s |
| $\dot{V}$ | : | volume flow rate, m$^3$/s |
| $\lambda$ | : | coefficient of friction in pipe |
| $\nu$ | : | coefficient of dynamic friction |
| $\rho$ | : | density, kg/m$^3$ |

## 1. Introduction
### 1.1. The GAPS project

The General AntiParticle Spectrometer (GAPS) experiment is an international project that addresses crucial questions of astrophysics, especially dark matter physics. GAPS will make highly sensitive observations of cosmic-ray antiparticles, such as antiprotons, antideuterons, and antihelium.[1–4] Various dark matter models have been advanced to predict the small but detectable number of antiparticles in the low-energy range around 100 MeV. The yet undiscovered antideuterons are of particular interest. The detection of such antiparticles would provide unique constraints for theoretical dark matter models.[5]

GAPS will provide the first high-statistics measurement of antiprotons and unprecedentedly sensitive search for antideuterons in the low-energy range around 100 MeV. To achieve high sensitivities for low-energy antiparticles, GAPS will deploy long-duration NASA balloon flights over Antarctica carrying a several-meter-scale instrument for multiple extended periods of about a month.

The GAPS instrument mainly consists of a central tracker and a two-layer time-of-flight (TOF) system that surrounds the tracker (Fig. 1). The tracker has over 1000 custom-built, lithium-drifted silicon [Si(Li)] detectors arranged in a three-dimensional matrix to identify the rare antiparticles. To identify antiparticles by measuring X-rays with a sufficient energy resolution, Si(Li) detectors must be cooled below about −40°C.[6–8]

### 1.2. Heat pipe and radiator to cool the detectors in flight

The Si(Li) detectors generate internal heat, $Q_1$, of about 0.1 kW. $Q_1$ and heat inputs from the surroundings should be removed to cool the detectors below −40°C. This is realized by a unique heat-pipe system (HPS) consisting of multi-loop capillary pipes.[9–12] The top of Fig. 2 shows the concept behind the design of the HPS. All the detectors are thermally coupled with the HPS. The detector heat is conducted to the heat pipes, absorbed by a two-phase working fluid in the pipe, then conveyed to the cooling section via an adiabatic section. The cooling section is coupled with a radiator attached to a payload sidewall. The radiator is designed to operate at −50°C or below during the flight, radiating heat to space in the anti-sun direction. Meeting this requirement will require a radiator with an unprecedentedly large area for balloon payloads (2.3 m$^H$ × 4 m$^W$), as shown in Fig. 1.

### 1.3. Necessity for cooling the detectors before the flights

Before the balloon flight, the functioning of the GAPS payload must be evaluated on the ground (in laboratories) in the U.S.A. and Antarctica before and after shipping, respectively. For the ground tests, the detectors and the radiator must be cooled below −40°C and −50°C, respectively, which are expected during the flight.

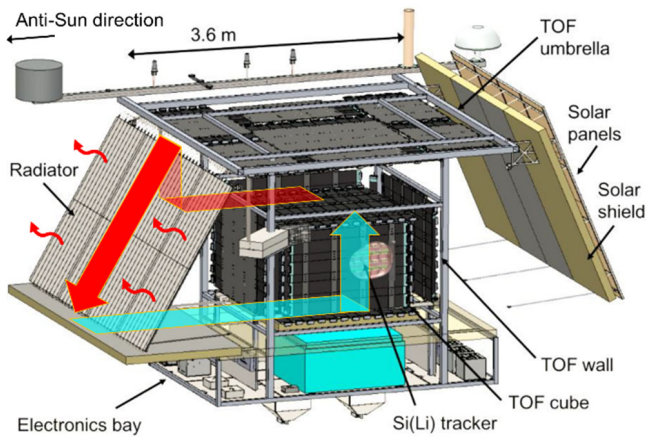

Fig. 1. Schematic view of the GAPS payload. On the radiator (4 m wide, 2.3 m high), 36 parallel heat pipes are arranged vertically and transfer heat from the tracker's detectors to the radiator. The radiator will be pointed away from the sun to radiate heat into space.

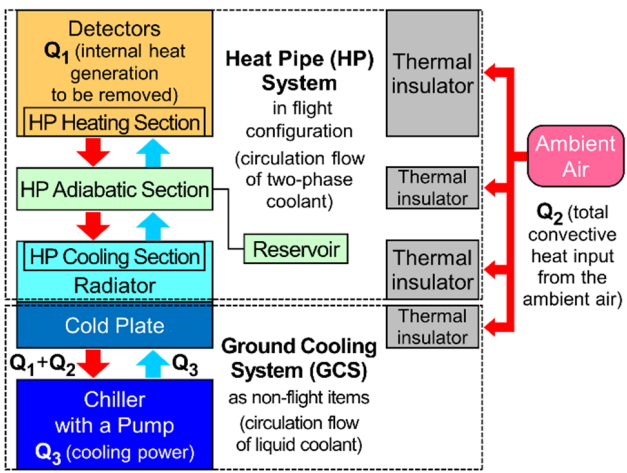

Fig. 2. Conceptual diagram of heat flow showing the heat-pipe system and the ground cooling system during ground tests. The internal heat generated, $Q_1$, and the heat input from the ambient air, $Q_2$, should equal the cooling power, $Q_3$.

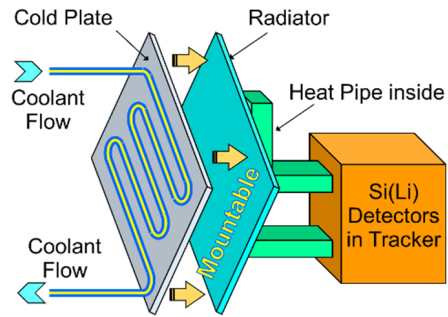

Fig. 3. Conceptual diagram of the ground cooling system. A cold plate is cooled by coolant and is attached to the radiator.

However, in a laboratory with the air at 1 atm and room temperature, the heat transfer between the radiator and the air is dominated by convection. Thus, the radiator cannot be cooled below −50°C by thermal radiation. Ground tests should be done in a laboratory without special facilities, such as thermal chambers, and without changing the payload flight configuration including the HPS.

Therefore, a simple approach must be established to cool the radiator easily.

### 1.4. The Ground Cooling System

This paper proposes a ground cooling system (GCS) to facilitate cooling tests before a flight. The GCS cools the radiator and has a simple design. The payload configuration is not changed for testing, but non-flight items may be attached to the payload.

Figure 3 shows the construction of the GCS. A cold plate is attached to the radiator to cool it. The cold plate is tightened to the radiator with bolts uniformly distributed in the plate so that the two surface planes make good contact. A tube is attached to the cold plate, and a coolant flows through the tube, drawing heat from the plate. Once the radiator is sufficiently cooled, the HPS transfers heat from the detectors to the radiator, bringing the detectors down to operating temperature.

### 1.5. Thermal insulation to realize the GCS

Assuming that the HPS, the radiator, and the detectors have an average temperature of −55°C, they are about $\Delta T \approx 80°C$ colder than the ambient air. In order to maintain this large temperature difference, the entire cooled apparatus must be thermally insulated from the ambient. Particularly, both the radiator and cold plate with a large surface area must be sufficiently insulated from a large heat input. Although the HPS itself is already thermally insulated in its flight configuration, its insulation thickness must also be confirmed to be enough for insulating in the laboratory.

In this study, we mainly use a standard thermal insulator of extruded polystyrene (EPS) foam to surround the cooling target. Despite the simple concept of this wrapping insulation approach, optimization of the insulator thickness is still a subject to date.[13] The thicker the insulator is, the larger the surface area becomes, resulting in larger heat input. Moreover, thicker insulator increases both the weight and cost which can be an issue for practical handling and manufacturing. Therefore, it is important to make a good estimation of heat input and to optimize the insulator thickness which effectively contributes to the thermal insulation.

Many research efforts have been carried out to optimize the thermal insulation for various applications in a wide dimension range, such as a scale of several tens of centimeters (e.g., cool boxes for keeping fishes fresh),[14] several meters (e.g., rockets with cryogenic tanks),[15,16] and several tens of meters (e.g., energy saving buildings).[17,18] The temperature difference between the outer and inner surfaces of insulator, $\Delta T$, is typically less than a few tens of degrees for buildings, $\approx 100°C$ for cool boxes using dry ice, and greater than 200°C for cryogenic apparatuses. Our GCS targets a combination of several-meter scale and $\Delta T \approx 80°C$. Hence, this study can provide unique insight to the practical research of thermal insulation.

## 2. Design concept of the GCS

Because of the payload's large surface area and the significant temperature difference between the radiator/heat pipes and room temperatures, heat input from the ambient air is considerable and must be removed. The heat input, $Q_2$, at thermal equilibrium can be estimated as follows:

$$Q_2 = \frac{S\, k\, \Delta T}{L}. \quad (1)$$

To roughly estimate $Q_2$, we assume that the HPS, the radiator, and the detectors have an average temperature of $-55°C$ ($\Delta T \approx 80°C$ colder than the insulator outer surface which must be close to the ambient air temperature) and are surrounded by extruded polystyrene (EPS) foam with an average thickness, thermal conductivity, and total surface area of $L \approx 0.1$ m, $k \approx 0.03$ W/m K, and $S \approx 40$ m$^2$, respectively. Based on these approximate estimations, Eq. (1) provides $Q_2 \approx 1$ kW. This 1 kW and $Q_1 \approx 0.1$ kW must be removed from the radiator.

### 2.1. Use of liquid nitrogen

In a case study of assuming pressurized liquid nitrogen (LN$_2$) flowed through the tube, the amount of heat removed by the LN$_2$ is expressed by:

$$Q = L_H \dot{m}. \quad (2)$$

Considering LN$_2$ has latent heat $L_H$ of ~200 kJ/kg, we found that more than $5.5 \times 10^{-3}$ kg/s of mass flow rate is necessary to remove 1.1 kW. Because the LN$_2$ density is ~800 kg/m$^3$, $\dot{m}$ represents an LN$_2$ consumption of ~0.6 m$^3$/day for 24 hours of continuous operation for a ground test, which will be run frequently. We consider that this exceeds the amount of LN$_2$, which is easily and continuously available in Antarctica.

### 2.2. The electric chiller

However, removing 1.1 kW of heat seems feasible if an electrical chiller is used. The bottom of Fig. 2 is a conceptual diagram of heat flow in the GCS using a chiller. The chiller circulates coolant fluid to remove heat. Sufficient thermal coupling of the cold plate and radiator ensures that heat is removed from the radiator. Many chillers with a cooling capacity larger than 1.1 kW are commercially available.

The idea behind the GCS is simple to implement, but the dimensions and the amount of heat to be removed are unprecedentedly large for pre-flight tests. Thus, we have developed the GCS in a step-by-step manner. Sections 3 and 4 of this paper describe our verification of its basic design by thermal testing of a scale model, and Section 5 discusses the details of the GCS development.

## 3. Basic study on cooling a plate by a chiller

A 1/10-scale GCS model was developed for testing, and the heat to be removed from the cold plate was scaled down to 100–200 W.

In the first step, a simple test was done to confirm that a scaled-down plate surrounded by insulators could be cooled by a chiller as expected.

### 3.1. Experimental configuration

Figure 4 shows the experimental configuration. A cold aluminum plate is uniformly encapsulated by 0.1 m thick insulation (DuPont Styrofoam Ace II), used as the EPS insulator. A copper tube with an inner diameter of 8.3 mm is embedded in the plate. A methacrylate adhesive (Plexus MA310) secures a strong thermal coupling between the copper tube and aluminum plate. It absorbs thermal stresses caused by the difference in thermal expansion coefficients of the two materials.[19] Heaters are attached to the cold plate to simulate the heat transferred by heat pipes to the plate.

The ends of the copper tube on the cold plate are extended to connect to a chiller that provides a circular coolant flow. A small chiller (Thomas TRL-70SLP) and a methanol coolant were used to evaluate this scale model. The tube is coiled around a reservoir downstream of the cold plate, which must also be cooled for the HPS to function. In this paper, we focus on cooling the cold plate and do not discuss details of how the HPS (including the reservoir) is cooled and operated.

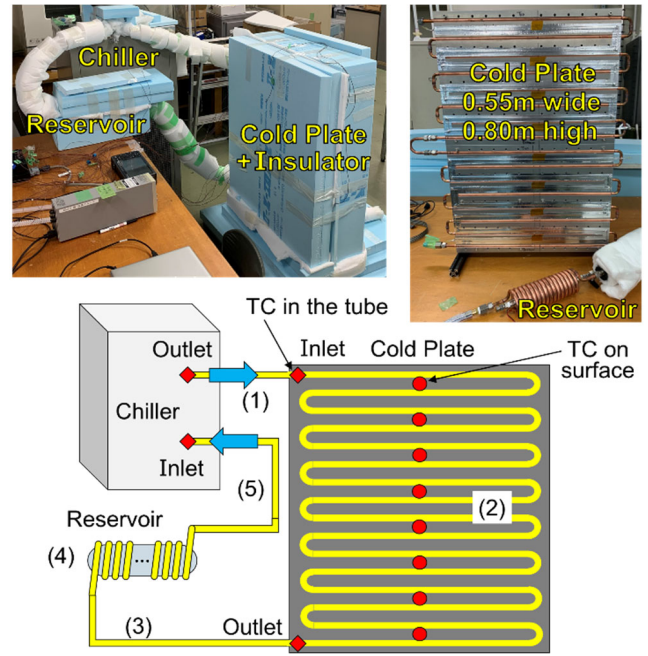

Fig. 4. Photos and a diagram of the experimental setup. A scaled-down cold plate is wrapped with insulation and cooled by a chiller unit. The temperatures at the red diamonds and circles are measured by thermocouples. The inner diameter and length of each tube section are summarized in Table 1.

Table 1. Experimental configurations for the 1/10-scale cold plate. Refer to the nomenclature list for each unit.

| Tube | (1) Flexible tube, chiller outlet | Inner $D$ = 0.0096, $L$ = 1.0 |
|---|---|---|
| | (2) Copper tube on cold plate | Inner $D$ = 0.0083, $L$ = 10.5 |
| | (3) Flexible tube from cold plate | Inner $D$ = 0.0120, $L$ = 1.0 |
| | (4) Copper tube on reservoir | Inner $D$ = 0.0079, $L$ = 2.9 |
| | (5) Flexible tube to chiller inlet | Inner $D$ = 0.0096, $L$ = 1.0 |
| Insulator | DuPont Styrofoam Ace II | $k$ = 0.028 (catalog value) |
| | Teijin/Bridgestone QonPET | $k$ = 0.041 (catalog value) |
| | Aeroflex MSR16 | $k$ = 0.031 (catalog value) |
| TC | Okazaki T35 | Sheathed, K type |
| | Ninomiya T-6F | T type |
| Chiller | Thomas TRL-70SLP | Cooling capacity = 250 W |
| | (Pump discharge $P$ = 0.03 MPa) | at lowest $T$ of −70°C |
| Logger | Graphtec GL820 | 1 Hz sampling |
| Heater | Clayborn A-16 | 12.5 Ω/m |

Sheathed thermocouples (TCs, K type) are mounted in the tube at the inlets and outlets of the cold plate and chiller. These TCs measure the methanol temperature directly (red diamonds in Fig. 4). The temperatures at various points on the cold plate are also monitored by TCs (T type) to confirm the temperature consistencies between methanol and the points on the cold plate (red circles in Fig. 4). A data logger records the temperatures at 1 Hz.

In addition to the cold plate, the tubes and reservoir are surrounded by 0.1 m thick insulation. QonPET non-woven fabric and Aeroflex elastomeric foam were wrapped around non-linear parts. Gaps between insulation were filled with a spray urethane foam (EA930TC-17) to minimize the heat input.

Design parameters and product names used for the experimental configuration are listed in Table 1.

### 3.2. Consistency between heat input and chiller cooling

The chiller outlet temperature was set at −70°C, and a heat load of $Q_1$ = 110 W was applied to the cold plate. Table 2 summarizes the results. The average temperature of the cold plate was −61.0°C. The insulator outer surface was 26.4°C on average, consistently with the room temperature. The total outer surface area of the insulator was 3.06 m². From Eq. (1), the heat input $Q_2$ was calculated to be 74.9 ± 1.7 W. The error is estimated from the uncertainty in measuring the temperature.

Here, we cross-check the heat input by using the temperatures of the methanol flowing in the tube. Assuming the flow is in quasi-equilibrium and is fully developed, the pressure loss of methanol flowing in the tube can be expressed by the Darcy-Weisbach equation as follows:

$$\Delta P = \lambda \frac{L}{D} \frac{\rho u^2}{2}. \quad (3)$$

Here, the coefficient of friction in the pipe, $\lambda$, is classified according to Reynolds number, or whether the flow regime is laminar or turbulent, by the following definitions:

$$\lambda = \begin{cases} 64\,Re^{-1}, & Re \leq 2300, \text{laminar flow,} \\ 0.3164\,Re^{-0.25}, & Re > 2300, \text{turbulent flow,} \end{cases} \quad (4)$$

$$Re = \frac{u\,D}{\nu}. \quad (5)$$

The amount of heat that methanol removes during its flow between the inlet and outlet of the cold plate is expressed by:

$$Q_3 = \Delta T\,\rho\,C_V\,\dot{V}, \quad (6)$$

$$\dot{V} = \pi\,(D/2)^2\,u = \dot{m}/\rho. \quad (7)$$

The pressure loss of methanol between outlet and inlet of the chiller, $\Delta P$, is fixed to 0.03 MPa, as defined by the discharge pressure of the chiller pump. For the coolant properties, we referred to Ref. (20).

The mass flow rate, $\dot{m}$, which is the same for every tube section, is calculated to be 0.042 kg/s. The calculated values of $Re$ are below 2000 throughout the chiller tube in this case. By combining the above equations and substituting the measured temperature difference $\Delta T$ between inlet and outlet of the cold plate, Eq. (6) gives the amount of heat removed from the cold plate by the coolant as $Q_3 \approx$ 183.8 ± 5.0 W. This is consistent with $Q_1 + Q_2 \approx$ 184.9 W, which is the amount of heat to be removed from the cold plate.

We successfully demonstrated that a scaled-down radiator surrounded by 0.1 m thick insulation could be cooled to −60°C in the laboratory. We also confirmed that the heat input to the cold plate is balanced with the heat removed from it by the coolant as well as that the heat leak can be suppressed to a negligible level by filling all gaps between insulations.

Table 2. Results of cooling a scaled-down cold plate by a small chiller.

| | |
|---|---|
| Cold plate surface temperature (averaged) | −61.0°C |
| Insulator outer surface temperature | 26.4°C |
| The outer surface area of cold plate insulator | 3.06 m² |
| Dummy heat load $Q_1$ | 110.0 W |
| Cold plate heat input $Q_2$ estimated from Eq. (1) | 74.9 ±1.7 W |
| Sum of $Q_1$ and $Q_2$ | 184.9 ±1.7 W |
| Chiller outlet temperature (setting) | −70.0°C |
| Methanol temperature at cold plate inlet | −63.9°C |
| Methanol temperature at cold plate outlet | −61.4°C |
| Chiller pump discharge pressure (fixed) | 0.3 MPa |
| Methanol mass flow rate | 0.042 kg/s |
| Heat removed from cold plate, $Q_3$, from Eq. (6) | 183.8 ±5.0 W |

## 4. 1/10-scale GCS model

### 4.1. Configuration to cool the scaled heat pipe system

In the next step, a 1/10-scale GCS model was developed to verify the GCS design. The scaled-down cold plate used in Section 3 was integrated into a scaled-down HPS to form the 1/10 scale-GCS model. The dummy heaters on the cold plate used in Section 3 were not used in this section. In the heating section of the HPS, 36 equivalent mock-up detector modules were mounted to simulate the heat dissipation from 144 Si(Li) detectors corresponding to 10% of detectors of the full-scale Si(Li) tracker.

Figures 5 and 6 show a conceptual diagram and photos of the scaled-down GCS model, respectively. Additional TCs were attached to various locations of the heat pipe system to confirm the heat pipe performance. Temperature data were taken under various conditions as summarized in Table 3; the chiller outlet temperature, reservoir temperature (or the detector target temperature of the HPS), and heat load applied to mock-up detector modules were parameterized. Case 1 is the typical instance. The other nine cases were used to study the heat balance under different conditions.

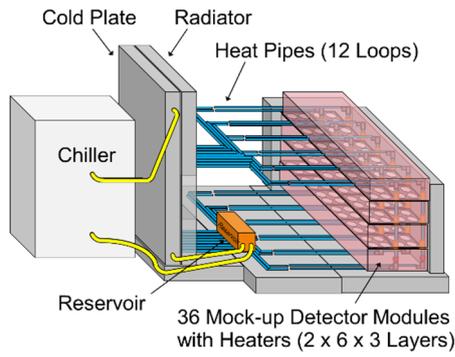

Fig. 5. Conceptual diagram of the 1/10-scale GCS.

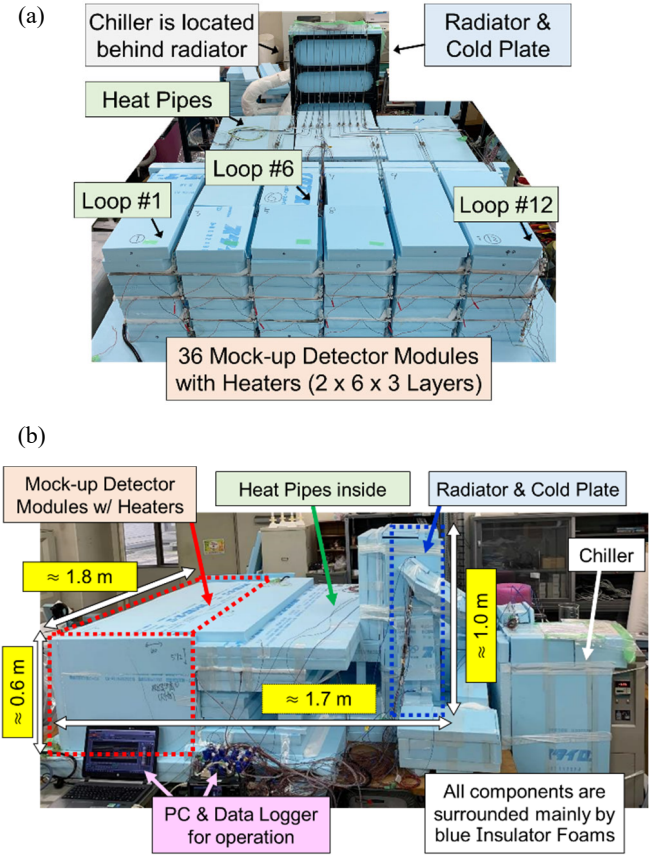

Fig. 6. The 1/10-scale GCS and HPS (a) during the integration and (b) after attaching all thermal insulators.

Table 3. Ten temperature combinations in the scaled GCS tests.

| | Chiller outlet $T$ | Reservoir $T$ | Total heat load on Modules | $\Delta T$ between detector and Ambient $T$ |
|---|---|---|---|---|
| Case 1 | −70°C | −51°C | 12 W | 67.9°C |
| Case 2 | −70°C | −46°C | 12 W | 63.9°C |
| Case 3 | −70°C | −21°C | 50 W | 36.3°C |
| Case 4 | −70°C | −21°C | 100 W | 28.2°C |
| Case 5 | −40°C | −26°C | 12 W | 44.5°C |
| Case 6 | −30°C | −6°C | 50 W | 21.9°C |
| Case 7 | −30°C | −6°C | 100 W | 14.8°C |
| Case 8 | −20°C | −6°C | 12 W | 29.1°C |
| Case 9 | −20°C | 4°C | 50 W | 11.3°C |
| Case 10 | −20°C | 4°C | 100 W | 4.8°C |

### 4.2. Cooling temperatures

Figure 7 shows an example temperature profile in the nominal case, Case 1. Although it took about 24 hours to reach equilibrium due to the limited cooling power of the chiller used for this study, all detector modules except three were successfully cooled below −40°C and remained stable, within ±1°C for another day. Even the three exceptions were cooled below −36°C, which is acceptable to serve for use in a health check. These three modules were at the edge of the module array, where heat input from the ambient air was larger. From these results, we have learned to make the thermal insulation thicker at the edge for the full-scale GAPS instrument to ensure that all detectors are cooled below −40°C.

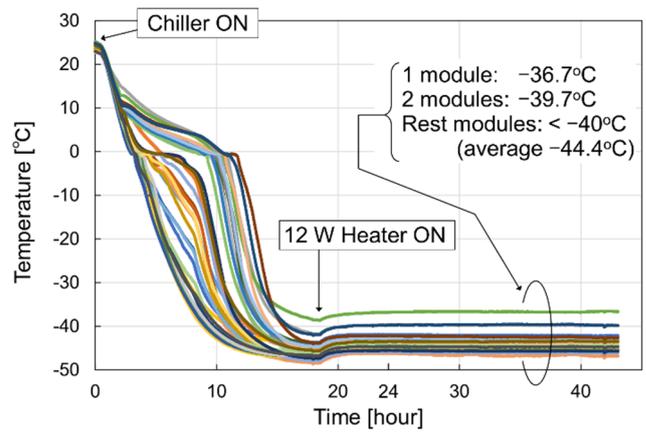

Fig. 7. Temperature profile for Case 1. Various color lines (36 lines) correspond to 36 detector modules. Within 24 hours, all detectors were cooled to equilibrium temperatures, below around −40°C.

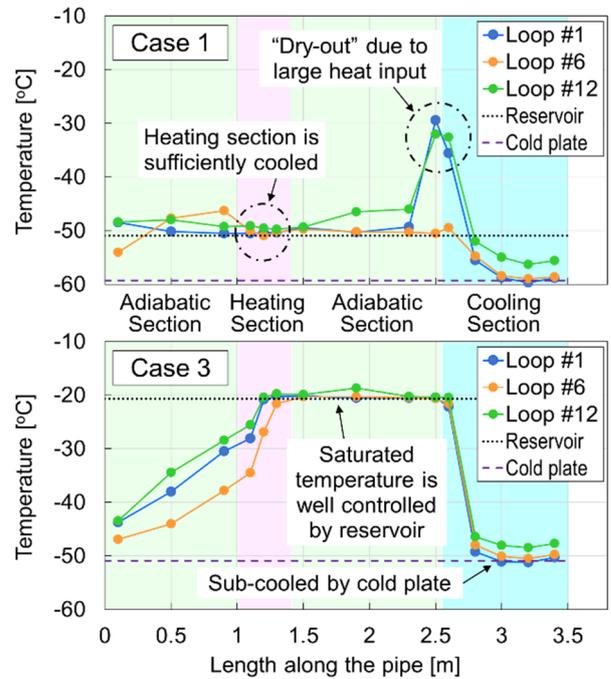

Fig. 8. Temperature profiles along three representative heat pipes loops (#1, #6, and #12) in Cases 1 and 3 are shown as examples. The scaled-down HP system functioned as expected. Locations of the three loops are shown in Fig. 6(a).

We believe that the temperature bumps around 0°C during cooling in Fig. 7 were caused by frost inside the insulation. Because the actual detectors will be packed in a dry bag filled with nitrogen gas (or dried air), these bumps are not expected in the full-scale case.

The cold plate temperatures showed non-uniformity less than 2.3°C which is sufficiently small for the HPS, especially in operation with a higher reservoir temperature. We confirmed that the heat pipe operated as expected. Although details about the heat-pipe data are omitted in this paper, temperature distributions along the heat pipes were as expected. As examples, temperature distributions for Cases 1 and 3 are shown in Fig. 8 (locations of TCs from each loop inlet, which differ among loops, are normalized to those for the first loop).

We have confirmed from these results that the scale HPS was successfully cooled and operated in the laboratory using the scaled-down GCS.

### 4.3. Heat balance

Here, we cross-check the heat balance of $Q_1 + Q_2 = Q_3$ at the cold plate similarly to the way we did in Section 3.2. Among these three terms, $Q_1$ is a given value, and $Q_3$ can be calculated from coolant temperatures, which are measured directly. Red triangles in Fig. 10 show $Q_3 - Q_1$ values estimated using Eqs. (3)–(7).

On the other hand, estimating $Q_2$ is non-trivial because the scaled HPS has a more complicated shape than the simple cuboid-shaped model used in Section 3, and it is surrounded by multiple, non-uniform insulation layers.

In this section, we propose to calculate $Q_2$ based on three different options and compare the calculated $Q_2$ values with

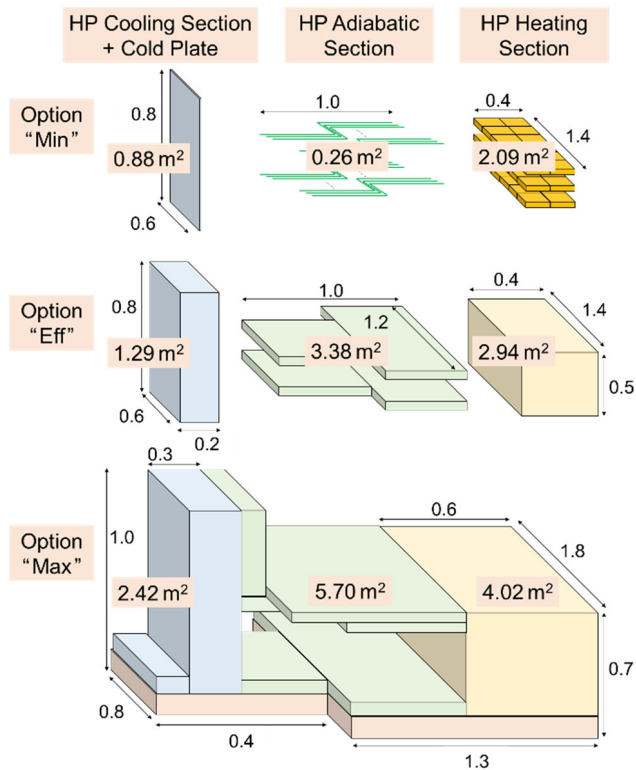

Fig. 9. Surface areas and approximate dimensions of cooling, adiabatic, and heating sections were used for the heat input estimation in three options Min, Eff, and Max.

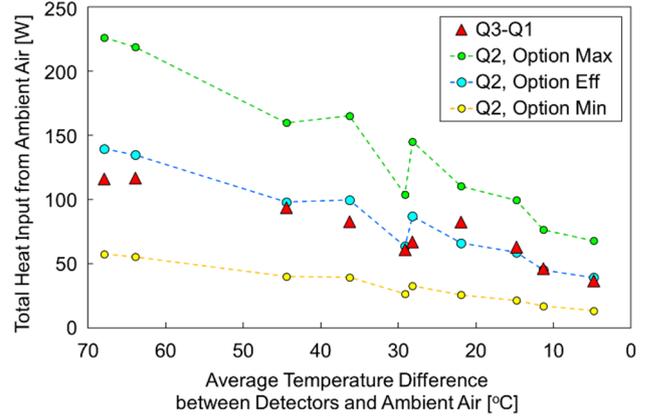

Fig. 10. Among $Q_2$ values estimated by three options, the Option "Eff" shows good agreement with measured $Q_3 - Q_1$ values for all ten conditions.

$Q_3 - Q_1$. The three options are as follows:

Option "Eff": the surface area, $S$, is defined by the outer surface area of the innermost insulators whose thicknesses are mostly 0.1 m (some range 0.05–0.1 m). Because results in Section 3 verified that these insulation thicknesses effectively wrap items that are ≈80°C colder than the ambient air, we predict this option will give us the effective area for Eq. (1).

Option "Min": the surface area is defined by the outer surface area of the "naked" scale model (without insulation). More precisely, $S$ is defined by the integrated inner product of area vectors of the outer surface area of the naked scale model and normal vectors of the inner surface of the innermost insulators. This option will provide extremely small $S$.

Option "Max": the surface area is defined by the outer surface area of the outermost insulation. The total thicknesses range in 0.1–0.2 m. Because insulators outside effective thicknesses won't majorly contribute to insulating, this option will provide extremely large $S$.

For these three options, $S$ can be calculated as shown by Fig. 9. Yellow, blue, and green circles in Fig. 10 show derived heat inputs for Min, Eff, and Max options, respectively. Blue circles show good agreement with red triangles. This indicates that insulation thickness of ≈ 0.1 m is sufficient to exclude heat from the GAPS detectors by the GCS. Moreover, it is verified that Option Eff is appropriate to estimate the effective surface area. It is handy to easily estimate the approximate amount of heat inputs from the ambient air without a detailed thermal analysis.

### 5. Full-scale Ground Cooling System

Based on the success of the scale model GCS, a full-scale GCS has been developed.

First, based on the Eff approach proposed in Section 4, the effective surface area of the full-scale GAPS instrument was calculated to be about 40 m². This is consistent with the approximate discussion in Section 2. Therefore, on the assumption of 0.1 m thick insulation, the heat input $Q_2$ from the surface area is estimated to be about 1 kW. The cold plate must remove this 1 kW together with $Q_1 \approx 0.1$ kW.

Next, the cold plate was designed to remove this 1.1 kW. In the scale model, a series of copper tubes to carry chiller coolant was designed to have a winding route on the cold plate (Fig. 4). The cold plate was thick enough to accept grooves to hold the tube to be embedded. This design has the advantage of cooling multiple-loop heat pipes uniformly. However, the design of the full-scale cold plate in this way has the disadvantage that the copper tube length needs to be very long due to the size of the full-scale radiator. Because a long tube increases coolant pressure loss, the higher discharge pressure is required for the chiller pump. Otherwise, the tube diameter should be larger to reduce the pressure loss, resulting in greater weight and harder handling. An evenly thick plate to embed the tube would also be excessively heavy.

Figure 11 shows the conceptual design of the full-scale GCS. The main upstream tube from the chiller is divided into four branches, each followed by a nine-split manifold. These 36 split tubes are mounted on the cold plate and then merged into the four downstream branches, followed by the downstream main tube connected with the chiller. On the cold plate, 36 split tubes are coupled with 36 aluminum strips. When the cold plate is mounted on the radiator, these strips are thermally coupled with the vertical 36 heat pipes on the radiator to ensure effective cooling. By this design, the weight of the cold plate is reduced to about 160 kg, which is feasible to handle in the laboratory. Ethanol will be used as the coolant instead of methanol for safety reasons.

By calculating Eqs. (3)–(7), the pressure loss, $\Delta P$, and volume flow rate, $\dot{V}$, of ethanol to absorb 1.1 kW can be estimated. Here, we assume that main, branch, and split tubes have inner diameters of 0.0111, 0.0238, and 0.0111 meters, respectively (given a standard tube-wall thickness of 0.8 mm). The temperature increase of ethanol between the outlet and inlet of the chiller is designed to be 5°C, which is acceptable for the HPS operation. As a result, $\Delta P$ and $\dot{V}$ must be larger than 0.06 MPa and 1.1×10⁻⁴ m³/s, respectively. A chiller with a cooling capability larger than 1.1 kW and meets the above requirements for $\Delta P$ and $\dot{V}$ can be used for the full-scale GCS. Considering these are very rough estimates, a chiller with excess capacity is desired. Among commercial products, for instance, the inTEST Thermonics A-80-1700 (cooling capacity = 1.7 kW at −80°C) and the Lneya LT-A050WN (capacity = 3 kW at −75°C) are two candidates.

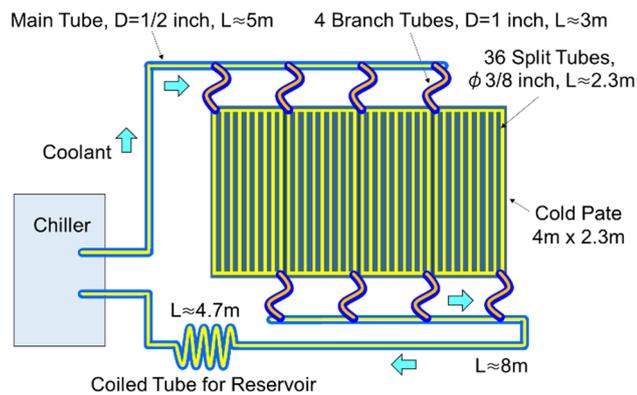

Fig. 11. Conceptual diagram of the full-scale GCS.

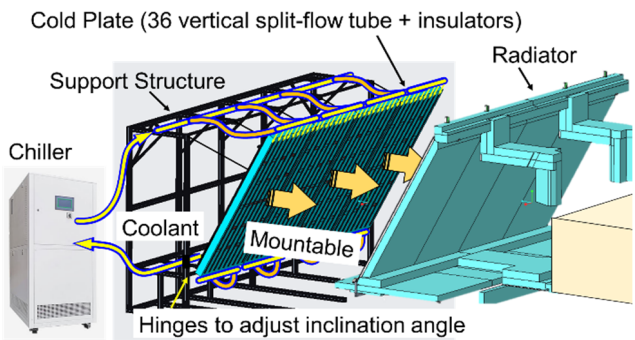

Fig. 12. The full-scale GCS is designed to be mountable/dismountable.

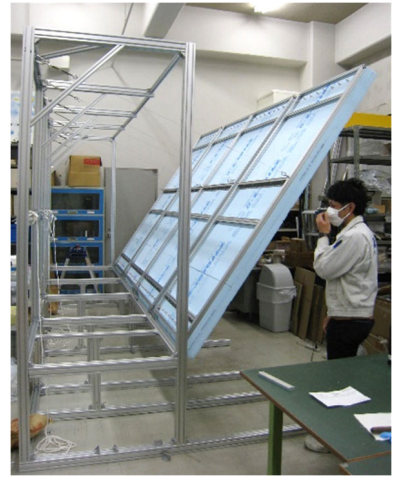

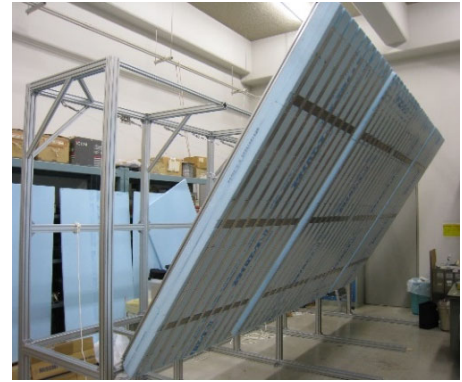

Fig. 13. Photos of the full-scale GCS temporarily integrated at ISAS/JAXA in Japan. Main tubes for the chiller coolant will be attached at the top and bottom of the tilted cold plate (2.3 m long, 4.0 m wide). The cold plate is filled with blue EPS foam insulators.

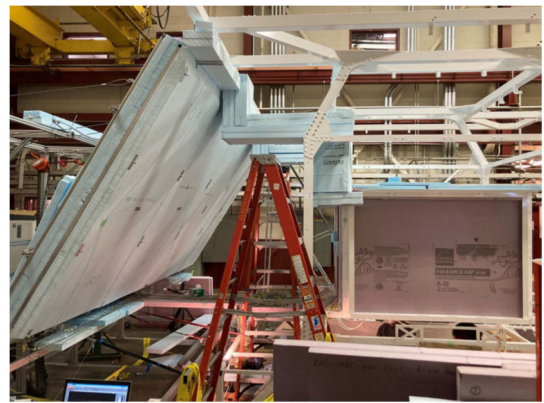

Fig. 14. The full-scale GCS and the GAPS payload being integrated at MIT in the U.S.A.

The cold plate is designed to be easily mounted and dismounted from the radiator. As shown in Fig. 12, the cold plate is designed to be supported without loading its weight onto the radiator.

The full-scale GCS was temporarily assembled in Japan to confirm mechanical connections between the components (Fig. 13). Then, the GCS was disassembled and transported to a collaborating research institute of Massachusetts Institute of Technology (MIT) in the U.S.A. As this manuscript is submitted in July 2022, integration of the full-scale GCS is in progress to be operated with the actual GAPS payload (Fig. 14). The details and thermal performance per ground tests in the U.S.A. and the technology transfer to the GAPS U.S.A. team to facilitate the GCS operation will be reported elsewhere.

## 6. Conclusion

The ground cooling system, GCS, was developed to realize the health-check tests of the GAPS balloon-borne payload on the ground before its flights. The GCS has a large cold plate wrapped by thermal insulators. The cold plate can be attached to the payload radiator without changing the payload flight configuration and is designed to cool detectors inside the payload via the HPS.

As far as we know, no previous research has developed such a cooling system with a meter-scale dimension, a large cooling temperature, and easily mountable/dismountable structure.

The cooling performance of the cold plate was investigated using a 1/10-scale GCS model consisting of a scaled-down cold plate and scaled-down HPS. Measured temperatures verified that the heat input from the ambient air is consistent with the heat removed by the chiller.

It was also confirmed that the heat input could be easily estimated from the surface area without using complicated thermal analyses or considering the air convection regime. The effective EPS thickness was found to be about 0.1 m for insulating a meter-scale instrument from the laboratory ambient which is 80°C higher than the cooling temperature. This thickness is comparable to (but a bit thicker than) typical values recommended for the building insulation (as expected by our larger $\Delta T$).[15,16] This simple and easy guideline will be useful to other applications including but not limited to aerospace and/or balloon experiments.

Based on the design verifications, we have developed the full-scale GCS whose integration is now in progress to be coupled with the GAPS payload.

**Acknowledgments**

This work is partly supported by JSPS KAKENHI grants JP17H01136 and JP19H05198 and Mitsubishi Foundation Research Grant 2019-10038. We express our sincere thanks to Naoyasu Suzuki, Shota Arakaki, Shunsuke Suzuki, Michael William Law, and all GAPS colleagues for their supports in developing and integrating the GCS.